\documentclass[entropy,article,accept,moreauthors,pdftex,10pt,a4paper]{mdpi} 

\usepackage{booktabs} 
\usepackage{multirow}
\usepackage{soul} 
\usepackage{microtype}

\setitemize{parsep=6pt,itemsep=0pt,leftmargin=*,labelsep=5.5mm}
\setenumerate{parsep=6pt,itemsep=0pt,leftmargin=*,labelsep=5.5mm}
\setlist[description]{itemsep=0mm}   
\setitemize{parsep=6pt,itemsep=0pt,leftmargin=*,labelsep=5.5mm,align=parleft}  
 \setenumerate{parsep=6pt,itemsep=0pt,leftmargin=*,labelsep=5.5mm,align=parleft} 

\firstpage{1} 
\articlenumber{390}
\doinum{10.3390/e20050390}
\pubvolume{20}
\pubyear{2018}
\copyrightyear{2018}
\history{}
\pdfoutput=1

 \theoremstyle{mdpi}
 \newcounter{thm}
 \newcounter{ex}
 \newcounter{re}
 \theoremstyle{mdpidefinition}

\Title{End-to-End Deep Neural Networks and Transfer Learning for Automatic Analysis of Nation-State~Malware}

\Author{Ishai Rosenberg *, Guillaume Sicard * and Eli (Omid) David *}
\AuthorNames{Ishai Rosenberg, Guillaume Sicard, and Eli (Omid) David}

\address[1]{Deep Instinct Ltd., Tel Aviv 6618356, Israel}

\corres{Correspondence: ishair@deepinstinct.com (I.R.); guillaumes@deepinstinct.com (G.S.);
david@deepinstinct.com (E.D.)}

\abstract{Malware allegedly developed by nation-states, also known as
\emph{advanced persistent threats}
(APT), are~becoming more common. The task of attributing an APT to a specific nation-state or classifying it to the correct APT family is challenging for several reasons. First, each nation-state has more than a single cyber unit that develops such malware, rendering traditional authorship attribution algorithms useless. Furthermore, the dataset of such available APTs is still extremely small. Finally, those APTs use state-of-the-art evasion techniques, making feature extraction challenging. 
In this paper, we use a deep neural network (DNN) as a classifier for nation-state APT attribution. We record the dynamic behavior of the APT when run in a sandbox and use it as raw input for the neural network, allowing the DNN to learn high level feature abstractions of the APTs itself. We also use the same raw features for APT family classification. Finally, we use the feature abstractions learned by the APT family classifier to solve the attribution problem. Using a test set of 1000 Chinese and Russian developed APTs, we achieved an accuracy rate of 98.6\%}


\usepackage[overlay,absolute]{textpos}

\begin{document}

\begin{textblock*}{10in}(55mm, 10mm)
{\textbf{Ref:} \emph{Entropy}, Vol.~20, No.~5, pp.~390--401, May 2018.}
\end{textblock*}

\section{Introduction}

While malware detection is always a challenging research
topic, a special challenge involves nation-state advanced persistent threats (APT), highly sophisticated and evasive malware. Since the usage of such cyber weapons might be considered an act of war~\cite{Hathaway12}, the question ``which country is responsible?'' becomes critical.

In this paper, we use raw features of dynamic analysis to train
a nation-state APT attribution classifier and to classify the APT family of each sample.

The main contributions of this paper are as follows:
\begin{enumerate}
\item	Implementing the first nation-state APT attribution classifier, which achieves a high accuracy on the largest test set of available nation-state developed APTs ever collected, successfully attributing new nation-state APT families.
\item	Providing the first nation-state APT family classifier.
\item	Showing that transfer learning is applicable for nation-state APT analysis, by using the high level feature abstractions of the APT family classifier for nation-state attribution as well.
\end{enumerate}

The rest of the article is structured as follows: Section~\ref{sec2} contains
the relevant related work to our use cases. Section~\ref{sec3} specifies the
problem definition and the unique challenges in this domain, both
with nation-state APT attribution and family classification in general, and especially when using
feature engineering. Section~\ref{sec4} describes our nation-state APT attribution and family classification implementation. Section~\ref{sec5} presents our experimental evaluation while Section~\ref{sec6} contains our concluding remarks and future work.

\section{Background and Related Work\label{sec2}}

There are numerous topics related to authorship attribution, such as plagiarism detection, books authorship attribution, source code authorship attribution and binary program authorship attribution. Stamatatos~\cite{Stamatatos09} provides
a broad review of many of those topics, including natural language processing (NLP) and traditional machine learning (ML) algorithms and the relevant features used, for example lexical (e.g., word frequencies), syntactic (e.g., sentence structure), semantic (e.g., synonyms) and application specific (such as a specific structure, for instance HTML).

In the following sub-sections, we focus only on the ones relevant to our work (and ignoring those that are irrelevant, such as source code authorship attribution, which cannot be used in our case since a source code is not available for the APTs).

\subsection{Binary Code Authorship Attribution\label{sec2.1}}

Certain stylistic features can survive the compilation process and
remain intact in binary code, which leads to the feasibility of authorship
attribution for binary code. Rosenblum et al.~\cite{Rosenblum11}
extracted syntax-based and semantic-based features using predefined
templates, such as idioms (sequences of three consecutive instructions),
$n$-grams, and graphlets. Machine learning techniques are then applied
to rank these features based on their relative correlations with authorship.

Alrabaee et al.~\cite{Alrabaee14} extracted a sequence of instructions
with specific semantics and constructed a graph based on register
manipulation, and applied a machine learning algorithm afterwards.

Caliskan et al.~\cite{Caliskan15} extracted syntactical features present
in source code from decompiled executable binary and applied their method to identify authors of malware in a leaked hacker forum dataset.

Though these approaches
represent a great effort in authorship attribution, it should be noted
that they were not applied to real malware. Furthermore, some limitations
could be observed including low accuracy in the case of multiple
authors, being potentially thwarted by light obfuscation and the
inability to decouple features related to functionality from those
related to authors styles. 

\subsection{Malware Attribution\label{sec2.2}}

The difficulty in obtaining ground truth labels for samples has led to
much work in this area to focus on clustering malware,
and the wide range of commonly used obfuscation techniques have led
many researchers to focus on dynamic analysis rather than using static features (i.e., instead of examining the raw file, focus on the report generated after running the file dynamically in a sandbox).

The work of Pfeffer et al.~\cite{Pfeffer12} examines information
obtained via both static and dynamic analysis of malware samples in order to
organize code samples into lineages indicative of the order in which
samples were derived from each other. 

Alrabaee et al.~\cite{Alrabaee17}
have used both features extracted from the disassembled malware code
(such~as idioms) and from the executable itself, and used mutual information
and information gain to rank them and built an SVM
classifier using
the top ranked features. 

The malware attribution papers mentioned so far require a large amount of pre-processing
and manual domain-specific feature engineering to obtain the relevant features.
Those methods are also applicable only to cases where a malware is an evolution of another malware (e.g., by mutation), or from
the same family, functionality-wise. These methods are not effective when completely different families of malware are examined. Our paper presents a novel application of DNN for APT attribution, specifying which nation has developed a specific APT, when the APTs in question are not derivatives of one another, and belong to completely different families.

The work of Marquis-Boire et al.~\cite{Boire15} examines several
static features intended to provide credible links between executable
malware binary produced by the same authors. However, many of these
features are specific to malware, such as command and control infrastructure
and data exfiltration methods, and the authors note that these features must
be extracted manually. To the best of our knowledge, this is the only
available paper that explicitly dealt with nation-state APTs detection
(using features common in them, such as APTs). However, those use
cases are limited, and no accuracy or other performance measures were provided. In addition, the paper did not deal with classifying which nation developed
the malware, and rather mentioned that one could use the similarities
between a known (labeled) nation-state APT to an unknown one to infer
the attribution of the latter.

\subsection{Malware and Malware Family Classification\label{sec2.3}}

Huang et al.~\cite{Huang16} used DNN as binary malware classifiers using both static features and dynamic ones. The usage of DNNs outperform state-of-the-art classic machine learning classifiers such as SVM, as specified by Hardy et al.~\cite{Hardy16}.

Dynamic analysis-based malware family classification using a DNN based on null-terminated patterns observed in the process’ memory, tri-grams of system API calls and distinct combinations of a single system API call and one input parameter was done by Huang et al.~\cite{Huang16} for state-of-the-art performance of 2.94\% error rate. Ahmadi et al.~\cite{Ahmadi15} presented a better accuracy (99.98\%) for a smaller data set (20,000 malware samples), using extensive feature engineering.

Huang et al.~\cite{Huang16} examined the concept of transfer learning for malware family classifications, using the layers trained for the binary classifier, except for the top layer.

However, none of those papers used nation-state APTs, which encompass unique challenges, detailed in Section~\ref{sec3.3}. Our research extends the usability of transfer learning to nation-state APT attribution, which is less trivial than malicious or benign functionality presented in~\cite{Huang16}, as specified in Section~\ref{sec5.1}. Furthermore, our paper uses raw features, making it more flexible and easier to generalize on future datasets, as we will see in the next section. 

Bengio et al.~\cite{Bengio13} discussed the usage of raw features and learning the feature hierarchy by DNNs, termed representation learning.

David et al.~\cite{David15} used words from Cuckoo Sandbox reports as features, similar to the way we use them in this paper, to generate malware signatures for malware family classification. However, both our our classifier's functionality (APT attribution) and implementation is different. Moreover, classifying APT families is more challenging then classifying malware families, due to the challenges mentioned in the following section.

In this paper we expand the work on use of deep neural networks for nation-state APT attribution in \cite{Rosenberg17}.

\section{Problem Definition\label{sec3}}\unskip

\subsection{Nation-State APT Attribution\label{sec3.1}}
Given an APT as an executable file, we would like to determine which nation
state developed it. This is a multi-class classification problem, i.e., one label per candidate nation-state.

\subsection{Nation-State APT Family Classification\label{sec3.2}}
Given an APT as an executable file, we would like to determine which family is it a part of APT families, like malware families, are group of samples with similar (malicious) behavior. When talking about APT families, such samples are usually being used as part of the same or similar campaigns. This is a multi-class classification problem, i.e., one label per nation-state APT family. Note that you can know the APT family without attributing it to a specific nation-state, and vice-versa.

\subsection{The Challenges of Nation-State Attribution and Family Classification\label{sec3.3}}

Trying to classify the nation that developed an APT can be an extremely
challenging task for several reasons that we detail here.

Each nation-state has usually more than a single cyber unit developing
such products, and there is more than a single developer in each unit.
This means that the accuracy of traditional authorship attribution
algorithms, which associates the author of source code or program
using stylistic features in the source code, or such features that
have survived the compilation, would be very limited.

These APTs also use state-of-the-art evasion techniques, such as anti-VM,
anti-debugging, code~obfuscation and encryption (\cite{Virvilis13}),
making feature extraction challenging. This challenge applies to both attribution and APT family classification.

Moreover, the number of such available APTs is small, since such APTs
tend to be targeted, used~for specific purposes (and, unlike common criminal
malware, not for monetary gain) and therefore are not available on
many computers. Their~evasion mechanisms make them hard to detect as well.
This results in a further decrease in the training set size from which
to learn.

Finally, since nation-states are aware that their APTs can be caught,
they commonly might try to fool the security researchers examining the APT
to think that another malware developer group (e.g.,~another nation) has developed it (e.g., by adding the APT strings in a foreign language, embedding data associated with a previously published malware, etc.). That is, unlike traditional authorship attribution problems, in this case the ``authors'' are actively trying to evade attribution and encourage false attribution.
The same is true for APT family classification: The APT developers want to prevent the blocking of the entire campaign when a single APT is being detected, and therefore, they try to prevent the association of two malware from the same family, by using different command and control (C\&C) 
servers, different malicious payloads, etc. This makes the classification of two APT from the same family a big challenge, compared to the malware family classification conducted by, e.g., David et al.~\cite{David15}.

Despite these issues, manual nation-state APT attribution and family classification is being performed,
mostly based on functional similarities, common strings, etc. For example,
the APTs Duqu, Flame and Gauss were attributed to the same origin
as Stuxnet following a very cumbersome advanced manual analysis (\cite{Bencsath12}). The question is: how can we overcome
these challenges and create an automated classifier (which does not require lengthy manual analysis)?

\subsection{Using Raw Features in DNN Classifications in the Cyber Security Domain\label{sec3.4}}

One of deep neural networks (DNN) greatest advantages is the ability
to use raw features as input, while learning higher level features
on its own during the training process. In this process, each~hidden
layer extracts higher level features from the previous layer, creating
a hierarchy of higher-level features.

This is the reason why deep learning classifiers perform better than traditional machine learning classifiers in complex tasks that requires domain-specific
features such as language understanding~\cite{Collobert11}, speech recognition, image recognition
\cite{Zeiler14}, etc. In such a framework the input is not high level features, which are derived manually based on limited dataset, thus not necessarily
fitting the task at hand. Instead, the input is raw features (pixels
in image processing, characters in NLP, etc.). The DNN learns a high-level
hierarchy of the features during the training phase. The deeper the
hidden layer is the higher the abstraction level of the features (higher-level
features).

While most previous work on applying machine learning to malware analysis relied on manually crafted features, David et al.~\cite{David15} trained DNN on raw dynamic analysis reports to generate malware signatures for use in a malware family classification context. In this paper, we similarly train a DNN on raw dynamic analysis reports but the goal is obtaining a different functionality (APT attribution rather than signature generation).

The benefits of using raw features are: 
\begin{enumerate}
\item	Cheaper and less time consuming
than manual feature engineering. This is especially true in the case
of nation-state APTs, where the code requires a lot of time to reverse
engineer in-order to gain insights about features, due to obfuscation
techniques commonly used by it, as mentioned~above. 

\item	Higher accuracy
of deep learning classifiers, since important features are never overlooked.
For~instance, in our nation-state APT attribution classifier, mentioned
in the next section, we~have used the technique suggested in~\cite{Olden02}
to assess the contribution of each of our features, by~multiplying
(and summing up) their weights in the network, where the highest value
indicates the most significant feature. We have seen that, besides
the expected API calls and IP strings of C\&C servers, arbitrary hexadecimal
values were surprisingly some of the most important features. A
security researcher might throw such addresses away, since they are
useless. However, those values were the size of data of specific PE
section which contained encrypted malicious shellcode, identifying
a specific APT family. 

\item	More flexibility due to the ability
to use the same features for different classification objectives.
For instance, our nation-state APT attribution classifier uses the
same raw features as our APT family classifier. Therefore, we can implement both using only a
single feature extraction process. This, in turn, facilitates the usage of transfer learning for training additional models, as we do for our APT attributer and family classifier in Section~\ref{sec4.4}.
\end{enumerate}

\section{Implementation\label{sec4}}

The challenges mentioned in the previous section require a novel approach
in-order to mitigate them. As mentioned before, the nation-state APT attribution problem
is not a regular authorship attribution problem, since more than a
single developer is likely to be involved, some of them might be replaced
in the middle of the development, etc. This makes regular authorship attribution
algorithms, using personal stylistic features irrelevant. Another
approach would be to consider all of the same nation-state APTs as
a part of a single malware family. The rationale is that common frameworks
and functionality should exist in different APTs from the same nation.
However, this is also not accurate: each nation might have several
cyber units, each with its own targets, frameworks, functionality (as shown in Section~\ref{sec4.2}),
etc. Thus, it would be more accurate to look at this classification
task as a malicious/benign classifier: each label might contain several
``families'' (benign web browsers, malicious ransomware, etc.) that
might have very little in common. Fortunately, DNN is known to excel
in such complex tasks.

This brings us to the usage of raw features: due to the benefits of this approach, mentioned in Section~\ref{sec3.4}, we would use raw features, letting the DNN build its feature abstraction
hierarchy itself, taking into account all available APT families.

\subsection{Raw Features Used\label{sec4.1}}

A sandbox analysis report of an executable file can provide a lot of useful
information, which can be leveraged for many different classification
tasks. In order to show the advantages of using raw features by a
DNN classifier, we have chosen raw features that would be used for different
classification tasks, e.g., both APT attribution and APT family classification.

Cuckoo Sandbox is a widely used open-source project for automated
dynamic malware analysis. It provides static analysis of the analyzed
file: PE header metadata, imports, exports, sections, etc. Therefore,
it can provide useful information even in the absence of dynamic analysis,
due to, e.g., anti-VM techniques used by nation-state APTs. Cuckoo
Sandbox also provides dynamic analysis and monitors the process system
calls, their arguments and their return value. Thus, it can provide
useful information to mitigate obfuscation techniques used by nation-state
APTs. We have decided to use Cuckoo Sandbox reports as raw data for
our classifiers due to their level of detail, configurability, and
popularity. We used Cuckoo Sandbox default configuration.

Our purpose was to let our classifiers learn the high-level abstraction
hierarchy on their own, without involving any manual or domain-specific knowledge. Thus, we used words only, which are basic raw features commonly used in the
text analysis domain. Although Cuckoo reports are in JSON file
format (JavaScript Object Notation), which can be parsed such that specific information is obtained from them, we did not perform any parsing. In other words, we treated the reports as raw text, completely ignoring the formatting, syntax, etc. Our goal was to let our classifiers
learn everything on their own, including JSON parsing, if necessary. Therefore,
the markup and tagged parts of the files were extracted as well. For
instance, in ``api: CreateFileW'' 
the terms extracted are ``api'' and ``CreateFileW'',
while completely ignoring what each part means.

Specifically, our
method follows the following simple steps to convert sandbox files
into fixed size inputs to the neural network:
\begin{enumerate}
\item	 Select as features the top 50,000 words with highest frequency in all Cuckoo reports, after~removing the words which appear in all files. The rationale is that words which appear in all files, and words which are very uncommon do not contain lots of useful information.

\item	Convert each sandbox file to a 50,000-sized bit string by checking whether each of the 50,000 words appear in it. That is, for each analyzed Cuckoo report, $feature{[}i{]}=1$ 
if the $i$-th most common word appears
in that cuckoo report, or 0 otherwise.
\end{enumerate}

In other words, we first defined which words participated in our dictionary
(analogous to the dictionaries used in NLP, which usually consist of the most frequent words in a language) and then we checked each
sample against the dictionary for the presence of each word, thus
producing a binary input vector. 

\subsection{Dataset\label{sec4.2}}

We trained a classifier based on Cuckoo reports of samples of APT
which were developed (allegedly) by nation-states. Due to the small
quantity of available samples, we used only two classes: Russia and
China (which are apparently the most prolific APT developers).

Our training-set included 1600 files from each class (training
set size of 3200 samples) of four APT families specified below from dozens of known campaigns of nation-developed APTs. We used 400 samples of each family. 400 files from the training set were used as a validation set (100 samples from each family). The test set contained additional 500 files from each class (test set size of 1000 files) from 2 different APT families, 500 files from each family. The labels (both ground-truth attribution and family) of all these files are based on well-documented and extended manual analyses within the cyber-security community, conducted during the past years. The APT family labels were taken from Kaspersky anti-virus.

Note that 
the above-mentioned separation between training and test sets completely separates between different APT families as well. That is, if an APT family is in test set, then all its variations are also in test set only. This makes the training challenge much more difficult (and more applicable to real-world), as in many cases inevitably we will be training on APT developed by one group of developers, and testing on APT developed by a completely different group.

\subsubsection{APT Families in the Training-Set\label{sec4.2.1}}
\begin{enumerate}
\item	\textbf{Net-Traveler}: Net-Traveler, 
AKA TravNet, is a keylogging and information gathering malware used in targeted attacks and for reconnaissance purposes against high-profile targets in a variety of sectors.
The malware has first been observed in 2004, and has been active up to late 2016.
It is attributed to a China ({\url{https://www.slideshare.net/RahulSasi2/netravler-apt-attributionplachina})}.

\item	\textbf{Winnti/PlugX}: PlugX is a Remote Administration malware (RAT) that has first been observed in 2012.
Used in targeted attacks against high-profile targets in various sectors, including government and military/aerospace targets.
PlugX is attributed to China ({\url{https://securityaffairs.co/wordpress/72208/apt/analyzing-winnti-umbrella.html})}.

\item	\textbf{Cosmic Duke}: Cosmic Duke, AKA TinyBaron, is an information stealing malware first observed in 2014.
Used in targeted attacks against government and military targets.
The malware’s main purpose is to exfiltrate files from infected machines, and is characterized by a very lightweight first stage.
It is attributed to Russia ({\url{https://www.f-secure.com/documents/996508/1030745/dukes_whitepaper.pdf})}.

\item	\textbf{Sofacy/APT28}: Sofacy Group, AKA APT28, Fancy Bear Crew, Tsar Crew, is a threat actor active since mid-2000’s targeting mainly the political sector.
The campaigns are characterized by a high level of sophistication and employ Spear-Phishing and use of Zero-Day exploits. It is attributed to Russia ({\url{https://www.alienvault.com/blogs/labs-research/from-russia-with-love-sofacy-sednit-apt28-is-in-town})}.
\end{enumerate}

\subsubsection{APT Families in the Test-Set\label{sec4.2.2}}
\begin{enumerate}
\item	\textbf{Derusbi}: Derusbi is a family of malware for both Windows and Linux, the main functionality of which is Remote Administration (RAT).
It was first observed in 2008.
Used in targeted attacks primarily against industrial sector and health care sector targets.
It is attributed to China ({\url{https://www.threatconnect.com/the-anthem-hack-all-roads-lead-to-china/})}.

\item	\textbf{Havex}: Havex is a Remote Administration malware (RAT) that has first been observed in 2013.
It was used in targeted attacks against industrial sector targets, and has also infected SCADA and ICS systems.
It is attributed to Russia ({\url{https://cyberx-labs.com/en/blog/dhsfbi-report-says-russian-cyber-units-attacked-critical-infrastructure-blackenergy/})}.
\end{enumerate}

In order to avoid turning the attribution task to an APT functionality classification task, we use the same functionality (RAT) for all families in the test set, regardless of the nation-state a family is attributed to. We know that our attribution classifier generalize properly, since in the training set only one nation-state (China) has RAT functionality, as shown in Section~\ref{sec4.2.1}.

\subsection{Network Architecture and Hyper-Parameters\label{sec4.3}}

Our DNN architecture is a 10-layers fully-connected neural network, with
50,000-2000-\linebreak1000-1000-1000-1000-1000-1000-500-2 neurons, (that is, 50,000 neurons on the input
layer, 2000 in the first hidden layer, etc.), with an additional
output softmax layer.

We used a dropout (\cite{Srivastava14}) rate
of 0.5 (ignoring 50\% of the neurons in hidden layers for each sample) 
and an input noise (zeroing) rate of 0.2 (ignoring 20\% of input neurons) to prevent overfitting. A ReLU (\cite{Glorot11}) activation function was used, and an initial learning rate of $10^{-2}$ which decayed to $10^{-5}$ over 1000 epochs. These hyper-parameters were optimized using the validation set.

The APT family classifier used exactly the same architecture and hyper-parameters, except that the output layer is of 4 neurons (since there were four families in the training-set; see Section~\ref{sec4.2.1}), instead of 2 (for 2 nation-states: Russia and China) in the attribution classifier.

\subsection{Applying Transfer Learning to Nation-State APT Classification\label{sec4.4}}
In the computer vision domain, it is common to use the feature abstractions learned for one classification task for another task, as done, e.g., by Donahue et al.~\cite{Donahue13}.

An interesting question is would such technique be applicable to nation-state APT analysis,~either?

To answer this question, we decided to train our APT family classification, which high-level features hierarchy should include high level representation of dynamic behavior, which differentiate between different APT families. We did the following:
\begin{enumerate}[labelsep=5mm]
\item Remove the top layer of 4 neurons
\item Freeze the weights of the remaining 9 layers.
\item Add a new 2 neurons layer.
\item Train only the last layer's weights with the attribution labels.
\end{enumerate}

If our hypothesis, that the same high-level feature abstractions can be used for both tasks, is~correct, this classifier would have an accuracy similar to that of the attribution classifier, with a much shorter training time (since we train only a single layer instead of 10).

\section{Experimental Evaluation\label{sec5}}\unskip

\subsection{APT Family Classification\label{sec5.1}}

We tested the accuracy of the APT Family Classification DNN model over the validation set. The~accuracy for the nation-state APT family classifier was \textbf{99.75\%}
on it.

Note that we could not use the test set, which contained only families that were not included in the training set, and therefore our classifier would not be effective against them.
However, as~mentioned in Section~\ref{sec3.3}, nation-state APT classification is more challenging than regular malware family classification: nation-state APTs trying to distance themselves from each other, so the capture of a single APT sample would not compromise other samples (or, e.g., a common C\&C server) and the blocking of a single campaign would not affect other campaigns.
Thus, for all state and purposes, taking separate samples from the validation has the same challenge as using non-nation-state malware families never seen before.

\subsection{Nation-State Attribution\label{sec5.2}}

We tested the accuracy of the Nation-State Attribution DNN model over the test set. The accuracy for the nation-state APT attribution classifier was \textbf{98.6\%} on the test set, which contained only families that were not in the training set.

These test accuracies are surprising in light of the complete separation of APT families between train and test sets. Inevitably in many cases the developers or even the developing units of the APT in train and test sets are different (e.g., APTs in train set developed by one Chinese cyber unit, and APTs in test set developed by another Chinese cyber unit).

Given this strict separation, and in light of the high accuracy results obtained, the results lead to the conclusion that each nation-state has (apparently) different sets of methodologies for developing APTs, such that two separate cyber units from nation A are still more similar to each other than to a cyber unit from nation B.

\subsection{Transfer Learning from APT Family Classification to Nation-State Attribution\label{sec5.3}}
We tested the accuracy of the new nation-state APT attribution classifier DNN model, after~replacing and retraining the APT family classifier top layer. The accuracy was \textbf{97.8\%} on the test set, which contained only families that were not in the training set.

\begin{figure}[H]
\centering
\includegraphics[width=0.8\textwidth]{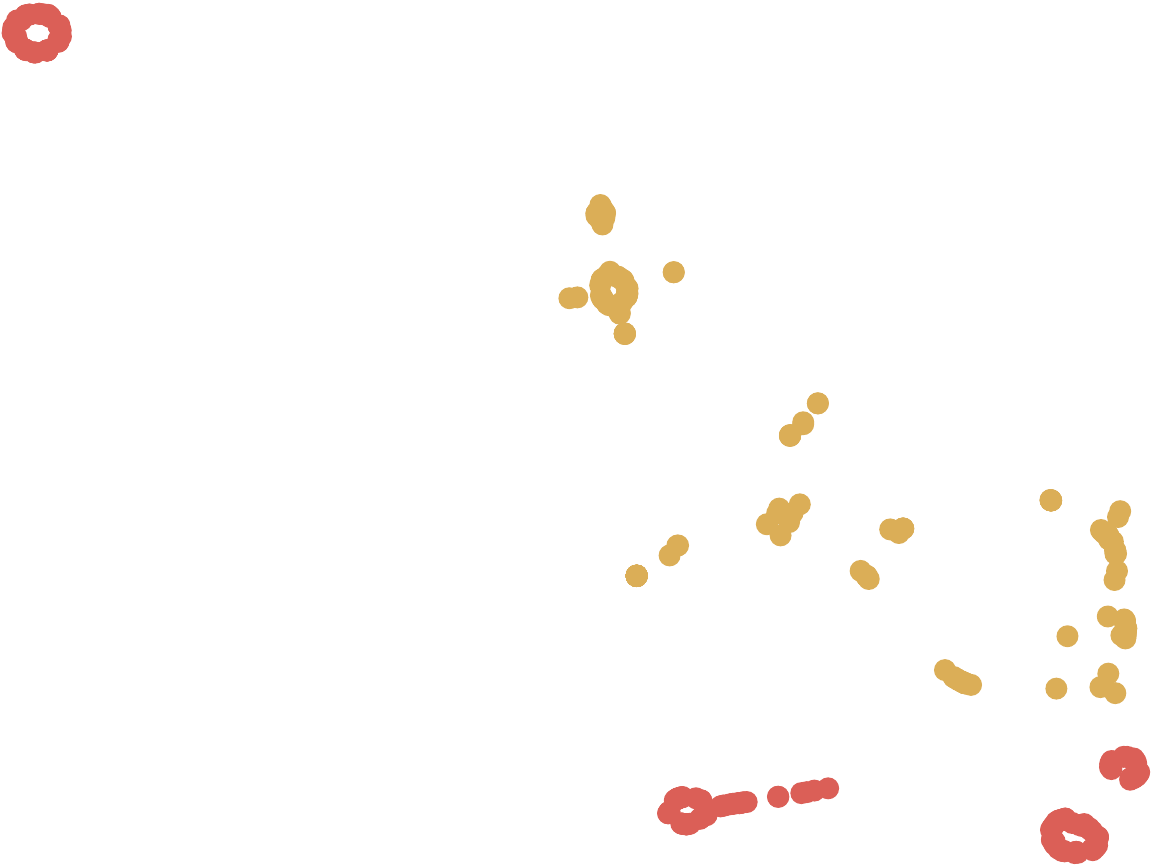}
\caption{A 2-dimensional visualization of the APT attribution (each node
is one nation-state), generated by the t-SNE dimensionality reduction
algorithm. Each color corresponds to one of two APT nation-states in the dataset. Note that
the labels are used for coloring the nodes only, and otherwise the visualization
is due to the DNN's learned features.\label{fig1}}\enlargethispage{0.5cm}
\end{figure}

\begin{figure}[H]
\centering
\includegraphics[width=0.8\textwidth]{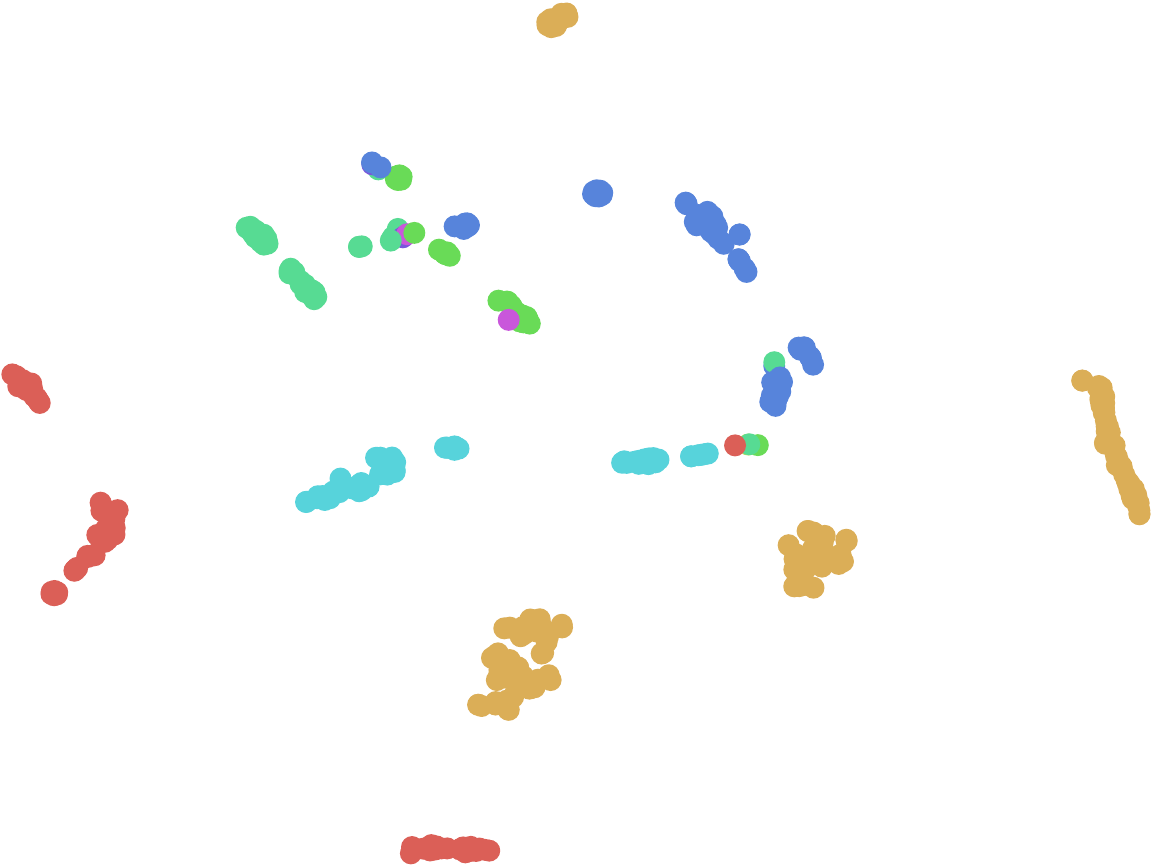}
\caption{A 2-dimensional visualization of the APT families (each node
is one APT family), generated by the t-SNE dimensionality reduction
algorithm. Each color corresponds to one of six APT families in the dataset. Note that
the labels are used for coloring the nodes only, and otherwise the visualization
is due to the DNN's learned features.\label{fig2}}
\end{figure}

To illustrate our results, Figures~\ref{fig1} and~\ref{fig2} provides a two dimensional visualization of the
data, where each node is one APT sample. The visualization
is generated using the t-distributed stochastic neighbor
embedding (t-SNE) algorithm~\cite{Maaten08}, in this case reducing the dimensionality of the data from 500 (number of neurons in the second-topmost layer) to 2 (number of nation-states in our dataset: Russia and China) and 6 (number of APT families in both training set and test set) respectably. The~goal of t-SNE is to reduce the dimensionality such that the
closer two nodes are to each other in the original high dimensional
space, the closer they would be in the 2-dimensional
space. Note~that the labels are used for coloring the nodes
only, and otherwise the visualization is due to the
DNN. The figure illustrates that variants of the same APT
family are mostly clustered together in the 500-dimension feature space,
demonstrating that the features learned by our DNN indeed capture
invariant representations of malware. Some clustering errors
are expected here (as can be seen in the visualization), since as
explained in Section~\ref{sec3.3}, many of these APTs try to distance themselves from other APTs from the same family by using features from other APT families, to thwart analysis and attribution attempts. Furthermore, as mentioned in Sections~\ref{sec4.2.1} and~\ref{sec4.2.2}, some APT families have different sub-families with slightly different behavior, so you often see several sub-clusters for the same APT family label (one same-colored sub-cluster for each sub-family). Here we use our ground truth labels, as mentioned in Section~\ref{sec4.2}, against
which we measure the performance of our method.

\unskip

At a first glance, the tasks of APT attribution and APT family classification seem entirely different: Attribution relates to implementing nation-state and not the functionality. Security researchers can analyze font types, C\&C servers locations, etc.
APT family classification, on the other hand, relates to the similarities in the functionality of the APT. Thus, you would expect a RAT from a specific country A to be more ``dynamic feature-wise similar'' to a RAT of country B, than to a key-logger of country A.
Thus, the such a high score for the attribution test set, using APT family classification feature abstraction seems odd.

However, further analysis shows that there are some subtle dynamic features or behavior that is common to all APTs from the same nation-state. For instance, specific code injection techniques are more commonly used by all APTs from a specific nation-state due to, e.g., more focus on those techniques in the training of the cyber unit of the regarded nation-state. The same is true for static features, either. For instance, the usage of the same development tools (e.g., a specific compiler version) by all the cyber units of the same nation-state, which might be different from the tools used by another nation-state, can affect the static features embedded in all the nation-state APTs (e.g., if Chinese file paths are longer, the debug section size might be different for the Chinese version of a compiler). 

Human security researchers could easily overlook such subtleties, being ``lost in the noise'' of possible features. However, as our results demonstrate, a DNN fed by raw feature can detect those subtle irregularities.

\section{Concluding Remarks\label{sec6}}

In this paper, we presented the first successful method for automatic APT attribution to nation-states, using raw dynamic analysis reports as input, and training a deep neural network for the attribution task. The use of raw features has the advantages of saving
costs and time involved in the manual training and analysis process. It also prevents
losing indicative data and classifier accuracy, and allows flexibility,
using the same raw features for many different classification tasks.
We also showed that the same raw features, and even the same high-level feature abstractions, can be used for other nation-state APT analysis tasks, such as APT family classification.

Our results presented here lead to the conclusion that despite all the efforts devoted by nation-states and their different cyber units in developing unattributable APTs, it is still possible to reach a rather accurate attribution. Additionally, different nation-states use different APT developing methodologies, such that the works of developers in separate cyber units are still sufficiently similar to each other that allow for attribution. The same conclusion can be drawn regarding APT families: despite the efforts to differentiate between APTs from the same family, they can still be accurately classified. The subtle dynamic feature differences between different nation-state cyber units even allows us to successfully use feature abstraction from on APT analysis task (family classification) in another (nation-state APT attribution).

While the work presented here could help facilitate automatic attribution of nation-state attacks, we are aware that nation-states could subvert the methods presented here such that they would modify their new APTs to lead to their misclassification and attribution to another nation-state. For~example, using deep neural networks themselves, they could employ generative adversarial networks (GAN)~\cite{Goodfellow14} to modify their APT until it successfully fools our classifier into attributing it to another nation-state. Applying GAN for APT modification would prove very difficult, but~theoretically possible. Another alternative is using adversarial examples that evade our classifier by adding ``benign features'', either dummy API calls or static features, as shown in~\cite{Rosenberg18}. 
Finally, as mentioned in Section~\ref{sec3.3}, since we use Cuckoo Sandbox to extract the features from the analyzed file, a nation-state APT can detect if it is running in a Cuckoo Sandbox environment and if so, quit immediately. However, this kind of behavior can also indicate an APT. Since we did not have a nation-state APT containing this functionality in our dataset, we could not asses its effect on our classifier, or whether the static features provided by Cuckoo Sandbox are enough to maintain accurate attribution even in this case.

In our future works in this area we will examine additional nation-state labels (multi-class classifier), once larger datasets of such APTs become available. We would also explore the usage of transfer learning between additional APT analysis tasks.

\vspace{6pt}

\bibliographystyle{mdpi}



\end{document}